\documentstyle[12pt,epsf]{article}
\newcommand{\be}{\begin{equation}}
\newcommand{\ee}{\end{equation}}
\newcommand{\bea}{\begin{eqnarray}}
\newcommand{\eea}{\end{eqnarray}}
\newcommand{\ci}{\cite}
\newcommand{\bi}{\bibitem}

\newcommand{\dd}{\partial}

\def\dal{\,\lower0.3ex\vbox{\hrule\hbox{\vrule\kern2pt\vbox{\kern4pt\kern4pt}
\kern2pt\vrule}\hrule}\,}

\begin{document}

\title{\sl Wave packet scattering from an attractive well}
\vspace{1 true cm}
\author{G. K\"albermann\\Soil and Water dept.,\\
Faculty of Agriculture, Hebrew University,
Rehovot 76100, Israel\thanks{e-mail:hope@vms.huji.ac.il}}

\maketitle

\begin{abstract}
\baselineskip 1.5 pc
Wave packet scattering off an attractive
well is investigated in two spatial dimensions numerically.
The results confirm what was found previously for the one dimensional case.
The wave scattered at large angles is a polychotomous (multiple peak)
coherent train.
Large angle scattering is extremely important for low impinging velocities and
at all impact parameters.
The effect disappears for packets more extended than the well.
Experiments to detect the polychotomous behavior are suggested.
\end{abstract}

\newpage
\baselineskip 1.5 pc
\section{\sl Introduction}

One dimensional wave packet scattering off an attractive potential well was
investigated in a previous work.\ci{pra}
In spite of being a thoroughly studied example of quantum
scattering for plane wave stationary states,
the effect found for packets, was unknown at that time.

Packets that are narrower than the well width initially,
recede from the well in the form of a multiple peaked
wave train.
Packets that are wider than the well, do not show this behavior.
A smooth wave hump proceeds both forwards and backwards.
Moreover, for narrow packets, the reflected waves dominate and scatter back
from the interaction region with an average speed that
is independent of the initial average speed of the
packet, whereas the transmitted waves proceed in accordance with
expectation.

Although wave packets seem to occupy a place of honor in the
educational literature of quantum mechanics\ci{merz}, they are
virtually absent from the research literature. Only fairly
recently, and mainly due to the arrival of bose traps
methods, there has been a resurgence of interest in the
subject.\ci{yeazell}
Conventional scattering processes are dealt with by using plane waves for the
incoming flux of particles. The justification for the approach,
stems from the fact that, accelerators generate beams
of particles that are almost monochromatic in energy (momentum),
and extremely spread in spatial extent.\ci{goldberger}
For such beams, the packets look more like a plane wave, and,
may be treated theoretically using stationary states.

The size of packets in actual scattering 
reactions is in any case enormous as compared to the
size of scatterers, therefore, the dependence on packet
details is irrelevant.

Some exceptions apply, however,
for atomic scattering processes, such as those investigated
in chemical physics\ci{tannor}. The semiclassical approximation is
used in the study of those processes to describe the actual motion of
individual atoms.

Present day capabilities of accelerators impede the
production of particle beams of spatial extent smaller
than the size of the scattering agents. Even optical
pulses in the femtosecond range are still wider
than the size of atoms from which they scatter.
It is nevertheless, not totally unrealistic, to
expect that the situation may change in the near future.
The import of the present paper reinforces the need
to produce narrow packets and design suitable
experiments.

The technique of cold bose traps may serve as such a setup,
because of the relative ease in handling beams of atoms at low
energy and their subsequent scattering inside cavities
taking the role of potentials.
In such an experiment with a narrow bunch of atoms, the scattered
atoms will proceed in a manner resembling the
coherent light emerging from a laser.

The {\sl ALAS} phenomenon in nuclear physics\ci{michel} may
also be related to the present findings, as described in section 4.

Polychotomous (multiple peak) waves are observed
when a superintense laser field focuses on an atom\ci{grobe}.
Ionization is hindered and the wave function is localized, in spite of
the presence of the strong radiation field.

In section 2 we will summarize and expand the results of the previous
work on the one-dimensional case.
Section 3 will describe two-dimensional scattering.
Some experiments are proposed in section 4.

\section{\sl One dimensional packet scattering off an attractive well}

In a previous work\ci{pra}, it was found that, a multiple peak coherent wave
train is reflected
from an attractive well, when the incoming packet is narrower than
the well.
These waves spend a large amount of time spreading out of the scattering
region.
The average
speed of the reflected wave was found to be independent of the
average energy of the packet.

The scattering was investigated in the framework of
 nonrelativistic potential scattering.
Narrow wave packets have high frequency modes.
One could suspect the approach not to be consistent, due to the emergence
of relativistic corrections for these modes.
However, we took precautions
by choosing a very large mass as compared to the 
inverse of the packet spread. We took {\sl m = 20}, while
the width of the packet was ${\Delta~x}^{-1}\approx{2}$ at least. 
For such a large mass, the speed of propagation of the 
frequency modes at the edge of the spectrum of the packet 
is still small in value, less than {\sl v=1} in our units.

Moreover, as will be depicted below, 
we are looking at an effect that unfolds immediately after
the packets starts to swell and not at very long times for which one
could doubt the validity of a nonrelativistic approach. 
The multiple peak behavior appears at times smaller than the spreading time 
of the packet. The relativistic corrections are then expected to
be of a lesser concern.

A correct treatment of high frequencies
or high momenta demands a relativistic wave equation.
Other wave equations must be considered in order to assess the
correctness of the above assumptions, such as the Klein-Gordon equation
or the Dirac equation. 
In spite of the limitations of the Galilean invariant Schr\"odinger 
equation, it has proven quite successful in atomic, molecular and 
nuclear processes, 
even for time dependent reactions, as the one dealt with presently.
This is the reason we opted for the nonrelativistic
potential scattering approach.

A narrow packet scatters backwards as a polychotomous
wave train, that is generated from the interference between the
incoming wave and the reflected wave.
For a narrow packet, the interference pattern is not blotted out
as time passes.
A very broad packet resembles more a plane wave, its spread
in momenta is much smaller than that of a narrow packet. 
When the well reflects waves in the backward
direction, they interfere destructively with the incoming broad packet, 
erasing the polychotomous behavior. 
A thin packet having short wavelength components
of the order of the well width (slit) produces a cleaner
diffraction pattern. Constructive interference
with the incoming packet, allows the pattern to survive.

Quantum mechanics textbooks show pictures of the development
of wave packet scattering from wells and barriers.\ci{schiff}.
Large oscillations of the wave function are seen when a packet is 
traversing a well. These oscillations are propagated
only in the backward direction. 

Only wide packets, as compared to the width of the well, 
are shown in the literature. The effect of the width of the packet 
is not investigated. What was found in ref.\ci{pra} and extended here,
is that the oscillatory behavior persists for narrow packets.

The question now arises as to the lifetime of the multiple peak
structure. Does it eventually die out and the peaks merge?
The answer to this question lies in the behavior of the
wave function inside the well.
We will show below that the scattering proceeds through 
a metastable, quasi-bound, state inside the well. This state 
does not decay exponentially, but polynomially in time. 
Differing from the transient behavior such as 
that of diffraction in time\ci{moshinsky}, for
which oscillations are set up 
by a shutter that is suddenly opened, 
the peak structure persists for very long times. Thousands of times
longer than the transit time of the packet through the interaction region.
Instead of  diffraction in time, we are witnessing a
diffraction in space and time.
From the numerical calculations, it appears that as far as we can observe
the long term behavior is still multiple peaked.\footnote{This aspect will be
addressed in a future work}

We now proceed to review the results of the one dimensional case
and add some further results. The next section will be devoted to
the two-dimensional case.

In ref.\ci{pra} we 
used a minimal uncertainty wave packet traveling from the left
with an average speed $v$, initial location $x_0$, mass $m$, wave number
$q~=~m~v$ and width $\Delta$,

\bea\label{packet}
\psi~=~C~exp\bigg({i~q~(x-x_0)-\frac{(x-x_0)^2}{4~\Delta^2}}\bigg)
\eea

The attractive well was located around the origin, with depth $V_0$ and width
parameter $w$.
We used an Gaussian potential, but
the results are not specific to this type of interaction.

\bea\label{well}
V(x)=-V_0~exp\bigg({-\frac{x^2}{w^2}}\bigg)
\eea
The packet above contains only positive energy components
and is therefore orthogonal to any bound state inside the well.
Any such superposition will be hindered by factors
of the form $e^{-\kappa |x_0|}$, where $\kappa=\sqrt{2~m~|E_B|}$,
 with $|E_B|$, the binding energy of the bound state. 
For initial locations $x_0$ faraway from the well, this
superposition vanishes.
However, the initial packet is not orthogonal to metastable 
states or quasi-bound states at positive energies.

We solved the Schr\"odinger equation for the scattering
event in coordinate space taking care of unitarity.
We used the method of Goldberg et al.\ci{gold}, that proved to
be extremely robust and conserves the normalization
of the wave function, with an error of less
than 0.01 \%, even after hundreds of thousands of time step iterations.
We also verified that the solutions actually solve the equation with
extreme accuracy by explicit substitution.

Figure 1 shows a series of pictures of the evolution of a narrow
wave packet.
The impinging packet has a width of $\Delta=0.5$, a momentum $q=1$,  and
the well width is $w=1$. We used a large mass $m=20$ in order
to prevent the packet from spreading too fast and to be on the safe
side regarding relativistic effects\ci{merz}.
\begin{figure}
\epsffile{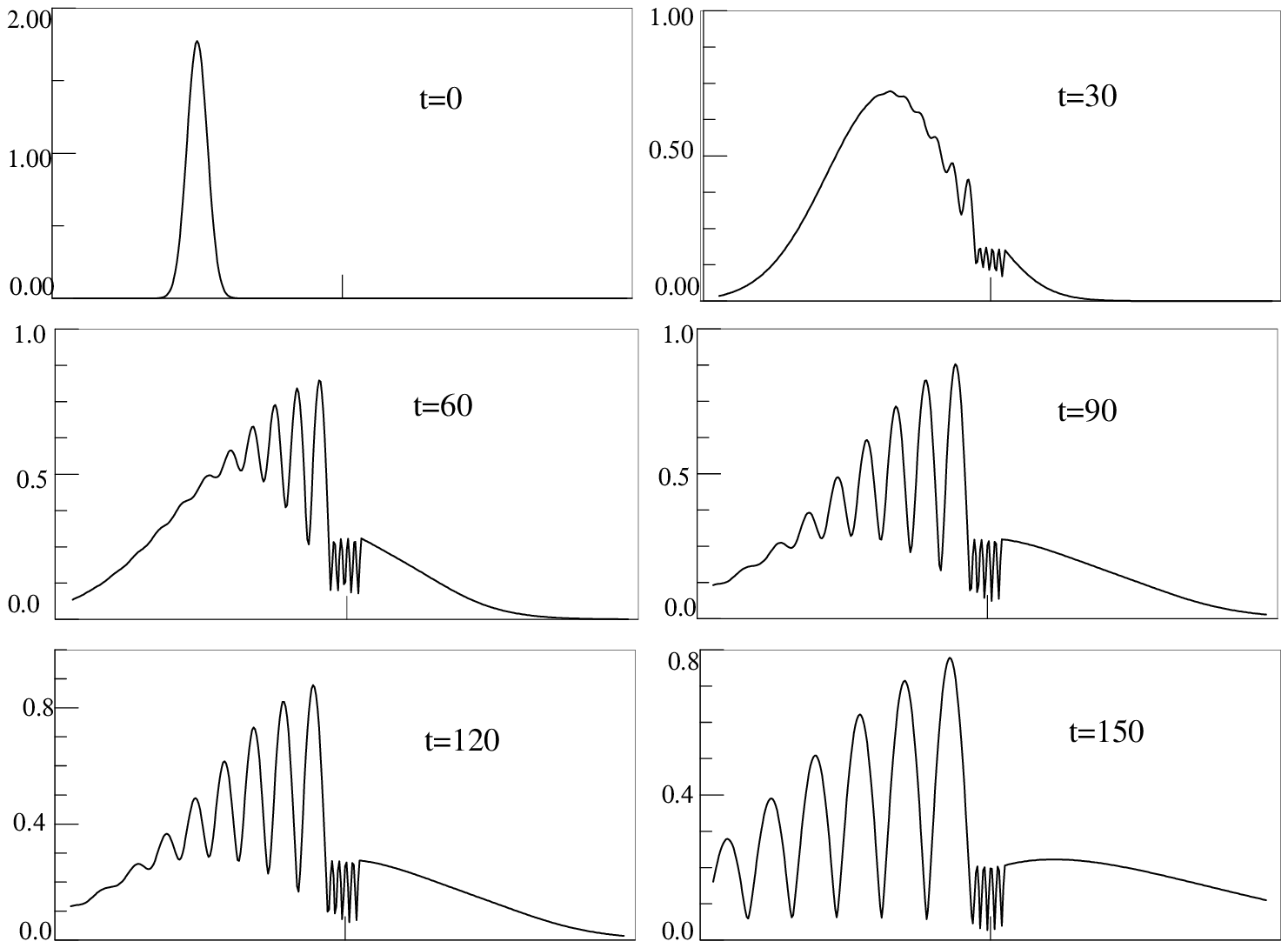}
\vsize=5 cm
\caption{\sl $|\Psi|$ as a function distance x for an initial
wave packet of width $\Delta=0.5$ starting at $x_0$=-10 impinging upon a well
of width parameter w=1 and depth $V_0$=1 at different times, the initial
average momentum of the packet is q=1.}
\label{fig1}
\end{figure}
The figure shows how the multiple peak structure
is produced early on. After $t=200$, the reflected wave train surpasses
the incoming wave and proceeds to propagate independently of it.
The effect persists for extremely long times.
Figure 2 depicts the scattered waves after t=5000, a time long enough for
the waves to scatter at a large distance (Recall that the well width is
w=1).
\begin{figure}
\epsffile{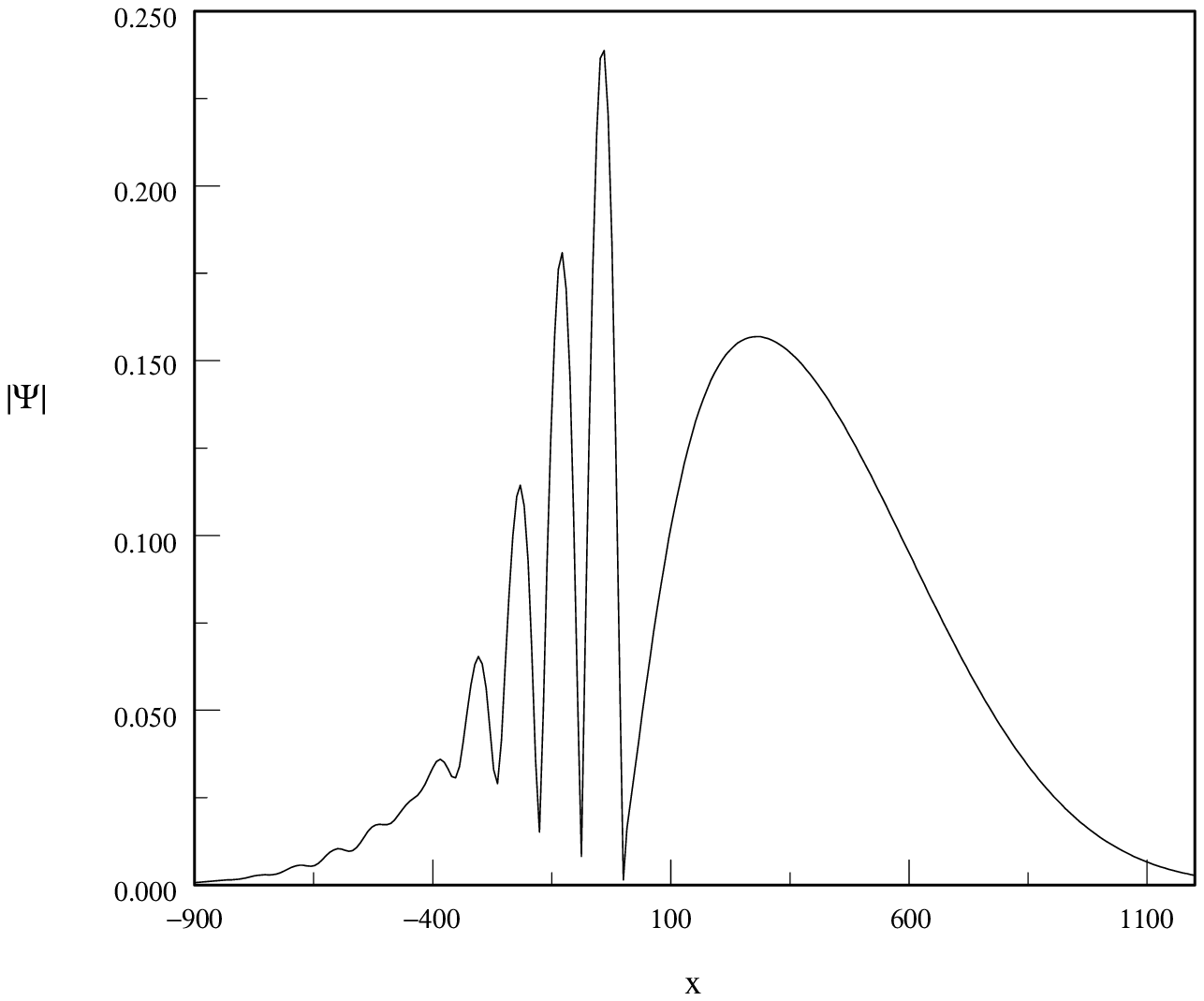}
\vsize=5 cm
\caption{\sl $|\Psi|$ as a function distance x for an initial
wave packet of width $\Delta=0.5$ starting at $x_0$=-10 impinging upon a well
of width parameter w=1 and depth $V_0$=1 after t=5000, the initial
average momentum of the packet is q=1.}
\label{fig2}
\end{figure}
A polychotomous (multiple peak) wave recedes from the well.
For low velocities, corresponding
to average packet energies less than half the well depth,
several peaks in the reflected wave show up.
Simple inspection reveals that the distance
between the peaks is constant.
The wave train propagates with an amplitude of the form

\bea\label{amplitude}
C(x)\approx e^{-\lambda |x|}~sin^2(kx)
\eea

The exponential drop is characteristic of a virtual
state solution inside the well.
The parameters $\lambda$ ad $\sl k$, are independent of the initial velocity,
but depend on time. The wave spreads and its amplitude diminishes, as expected.
We corroborated that the polychotomous behavior continues for
as long time as we could check numerically.

Figure 3 shows the sequence of events for
a wide packet. 
\begin{figure}
\epsffile{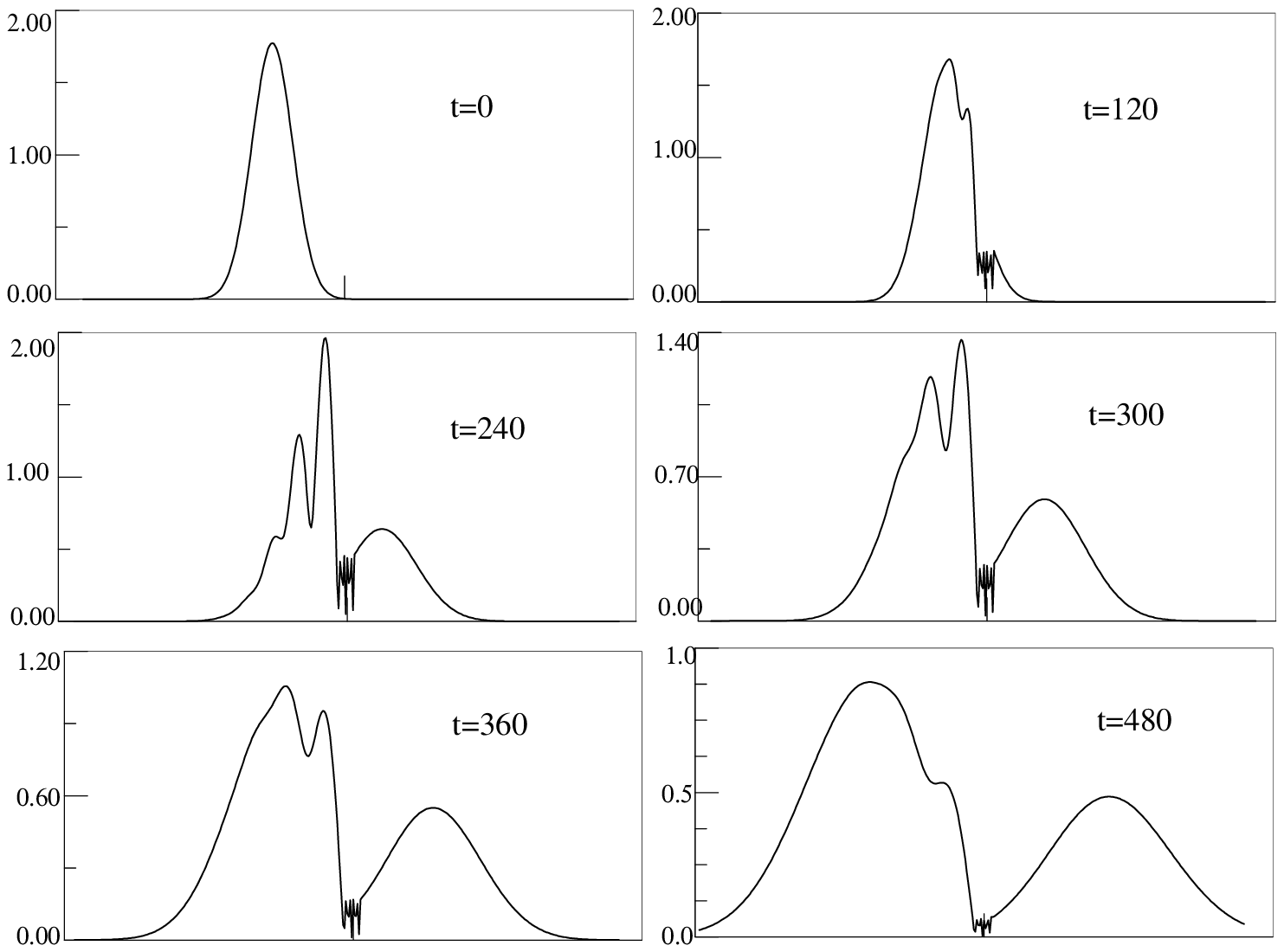}
\vsize=5 cm
\caption{\sl $|\Psi|$ as a function distance x for an initial
wave packet of width $\Delta=2.$ starting at $x_0$=-10 impinging upon a well
of width parameter w=0.5 and depth $V_0$=1 at various times, the initial
average momentum of the packet is q=1.}
\label{fig3}
\end{figure}

The multiple peaks disappear completely for packets wider than the well.
The long time behavior of the same case is depicted in figure 4.
\begin{figure}
\epsffile{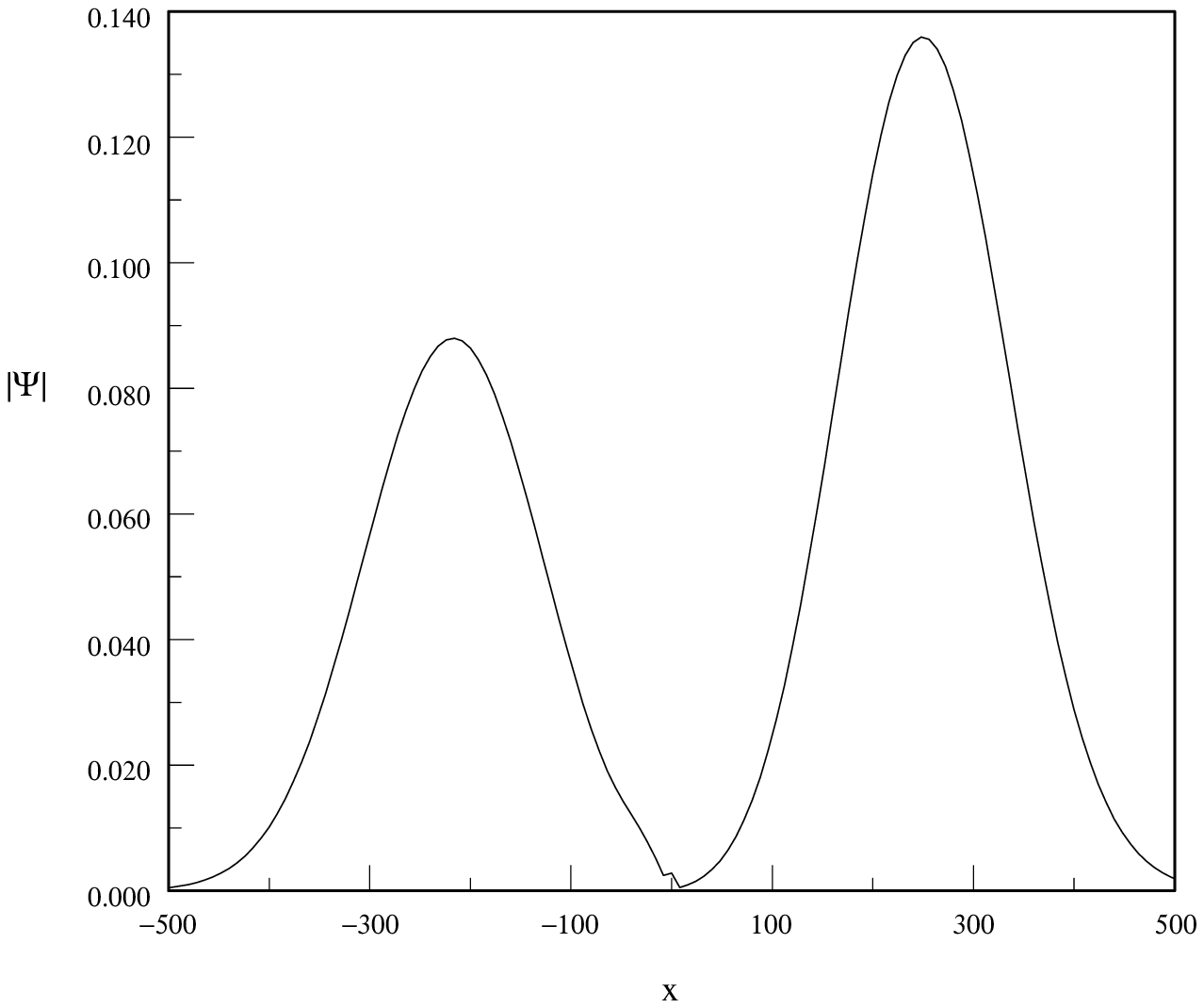}
\vsize=5 cm
\caption{\sl $|\Psi|$ as a function distance x for an initial
wave packet of width $\Delta=2$ starting at $x_0$=-10 impinging upon a well
of width parameter w=0.5 and depth $V_0$=1 after t=5000, the initial
average momentum of the packet is q=1.}
\label{fig4}
\end{figure}
An approximate expression for the multiple peak reflected packet
average speed was found to be, $v=k(t_{formation})/m$. Where $k$ represents
the wavenumber outside the well at the time it starts emerging from it
after a long period of multiple reflections.
This speed was found to be independent of the initial
packet speed. The memory of the initial packet is deleted.

We investigated other types of potentials, such as a
Lorentzian, a square well, etc.,
and found the same phenomena described here.
Moreover the effect is independent of the shape of the packet
as long as it is narrower than the well width.
We used square packets, Lorentzian
packets, exponential packets, etc., with analogous results.
\begin{figure}
\epsffile{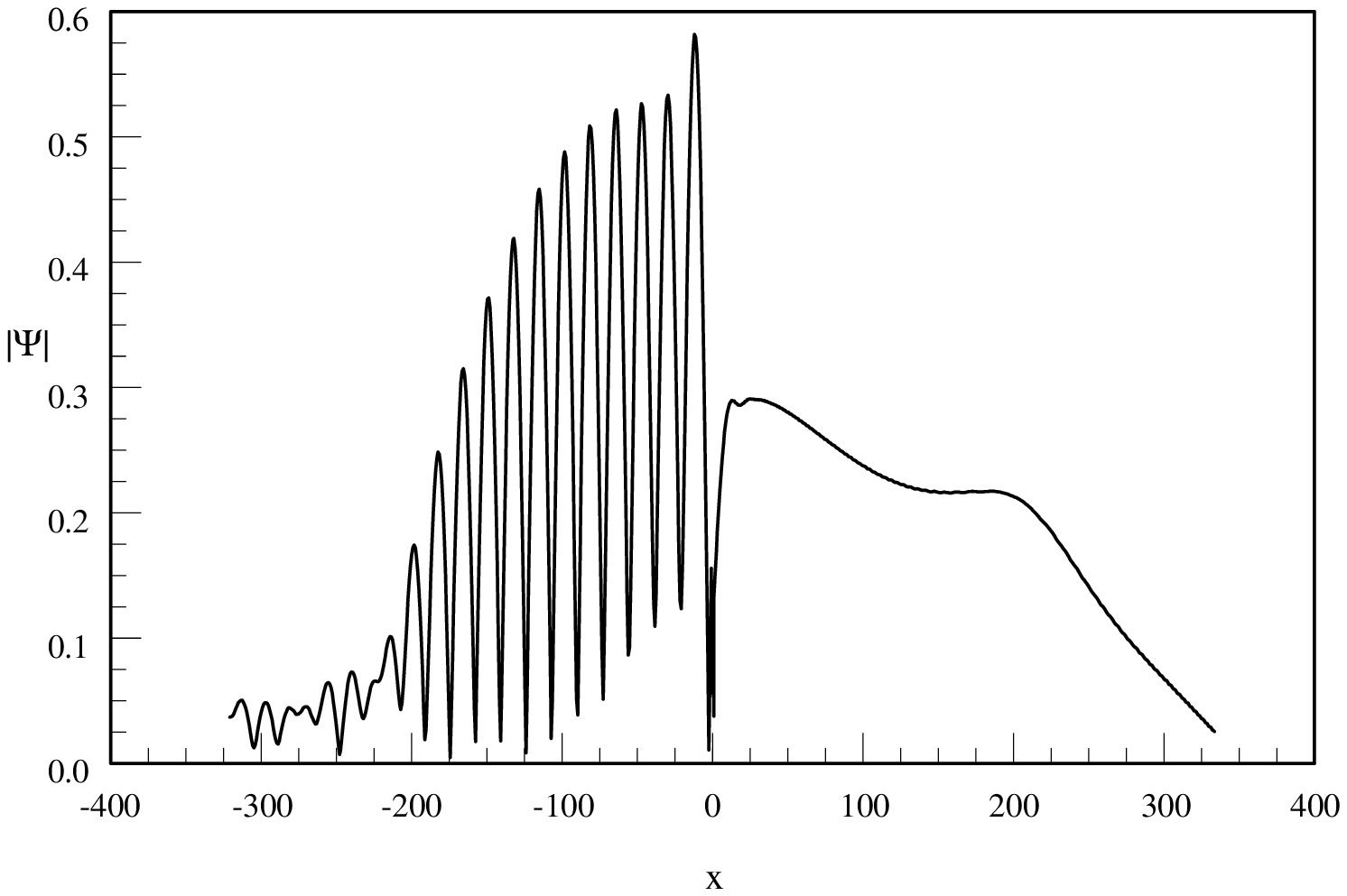}
\vsize=5 cm
\caption{\sl Theoretical calculation for a square initial packet
scattering off a square well for t=1000}
\label{fig5}
\end{figure}

In order to find analytical support, we resorted to
a square packet

\bea\label{square}
\psi(x)=e^{i~q~(x-x_0)}~\Theta(d-|x-x_0|)
\eea

where $d$ is half the width of the packet, $x_0$ the initial position
and $q$ the wave number.
It impinges on a square well located at the origin, whose
width is $2a$ and depth $V_0$

\bea\label{sqwell}
V(x)= -V_0~\Theta(a-|x|)
\eea
This case is solvable using the techniques of ref.\ci{zidell}.
The method is appropriate only for packets with sharp edges, that
terminate at a certain point. It consists of integrating the Fourier amplitude
of the wave using a contour in the complex momentum plane
that avoids the poles of the scattering matrix corresponding to the
bound states. For each momentum, one uses the appropriate stationary
scattering state for the square well.
The integral reads
\bea\label{integral}
\psi(x,t)=~\int_{\sl C}\phi(x,p)~a(p,q)~dp
\eea
where $\sl C$ is a contour that goes from $-\infty$ to $+\infty$
and circumvents the poles that are on the imaginary axis
for $P~<~i\sqrt{2 m~V_o}$ by closing it above them.
$\Phi(x,p)$ is the stationary solution to the square well scattering
problem for each $\sl p$ and $\sl a(p,q)$ is the Fourier transform
amplitude for the initial wave function with average momentum $\sl q$.
The results are depicted in figure 5.
The initial wave packet had average momentum $q=1$
width $\Delta=0.5$, and the square well parameters were $V_0=1,~a=1$.
The reflected wave shows exactly the same
polychotomous behavior as the numerical simulations.
In particular, numerical calculations with a square packet and a
square well match almost exactly the analytical results.

The polychotomous
effect is general, even the packet amplitude becomes unimportant.
The very existence of the effect does not depend
on the initial position of the packet, as mentioned in
passing in \ci{pra}.
Figure 6 shows one such case for the same parameters
as those of figure 1, but an initial location of $x_0=-50$.
The number of peaks has increased
and the distance between them has shrunk.
There appears a smooth
background under the multiple peaks. The well reacts to the
presence of the packet from far away. 
So, even if the packet is narrower only faraway from the well, the
polychtomous structure persists, despite the normal spreading
that must occur until the center of the packet reaches
the well, which, in the depicted case, would amount to
many times the original width. 
\begin{figure}
\epsffile{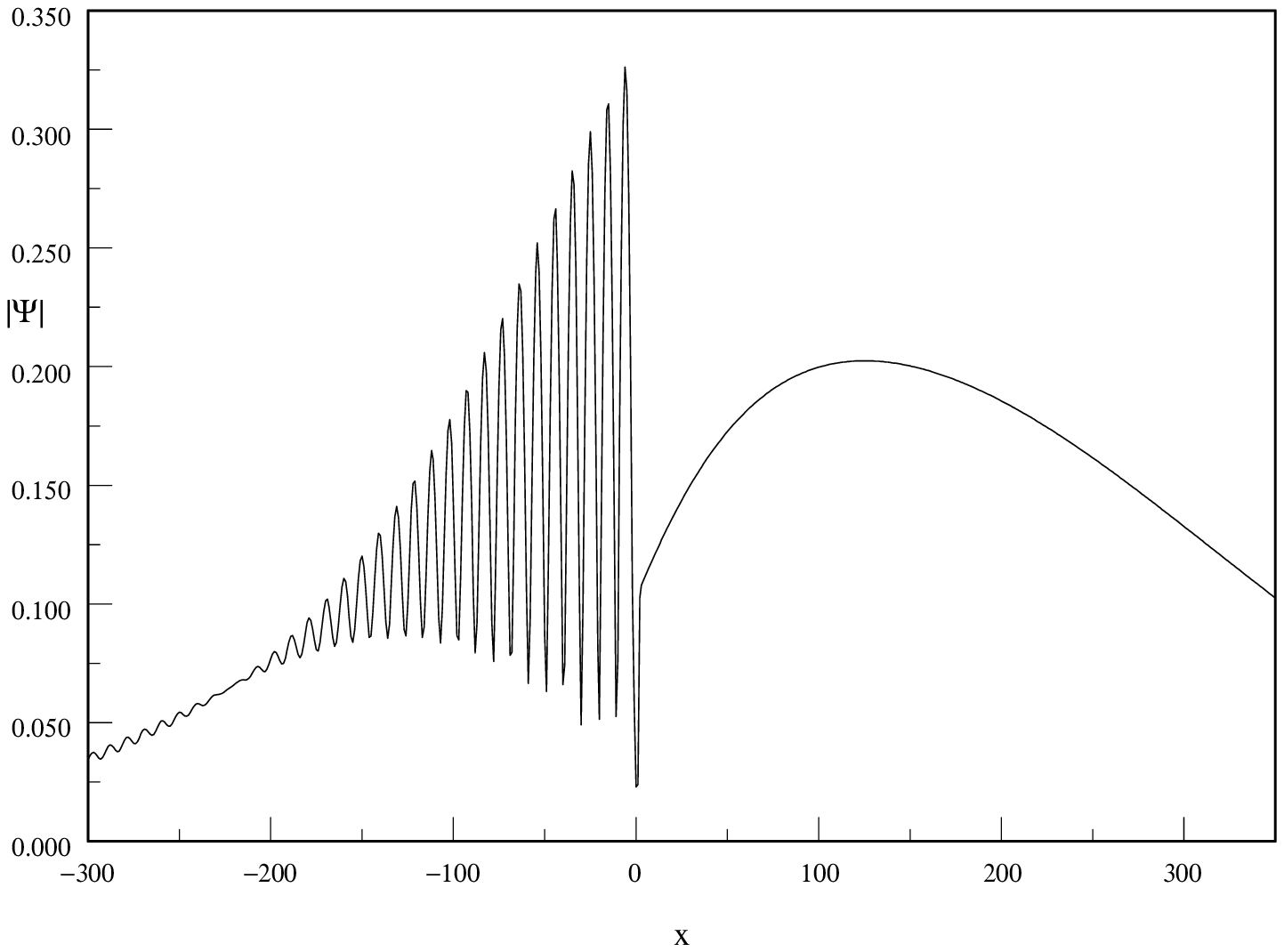}
\vsize=5 cm
\caption{\sl $|\Psi|$ as a function distance x for an initial
wave packet of width $\Delta=0.5$ starting at $x_0$=-50 impinging upon a well
of width parameter w=1 and depth $V_0$=1 after t=3000, the initial
average momentum of the packet is q=1.}
\label{fig6}
\end{figure}

It was claimed above that the multiple peak effect, was due
to the interference between incoming and reflected waves.
A sign of the its persistence may be found in the
behavior of the wave inside the well.
The amplitude of the wave function at the
origin as a function of time provides us with a suitable
index to characterize it. 
The long time behavior of the decay of this amplitude 
is polynomial. Trial and error lead us to a fit with
a polynomial of the form $|\psi(0)|=\frac{C}{t^{1.55}}$, with {\sl c},
a constant. Figure 7 shows this time dependence for the case of a narrow 
packet.
\begin{figure}
\epsffile{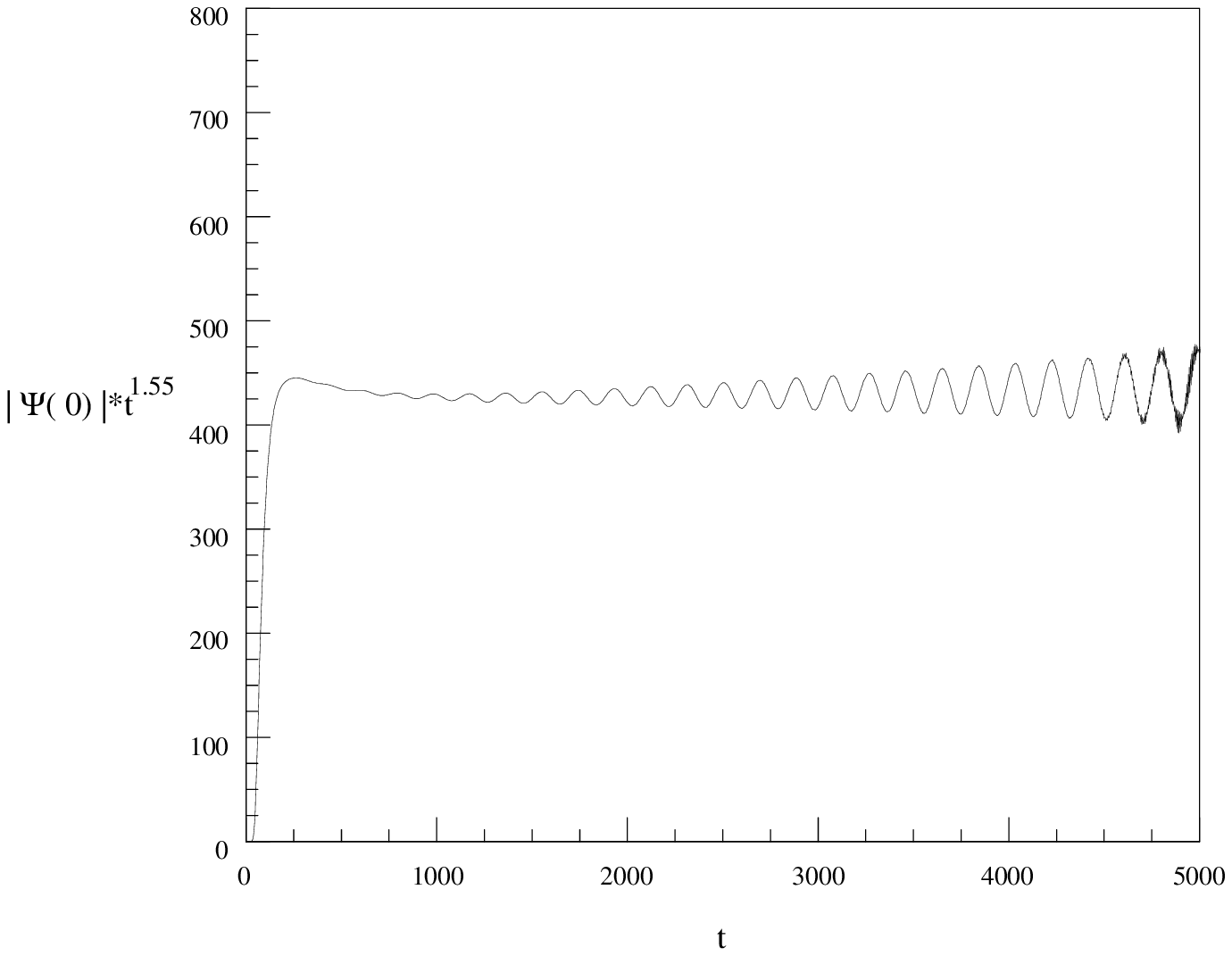}
\vsize=5 cm
\caption{\sl $|\psi(0)|$ as a function of time for the scattering depicted in 
figure 1}
\label{fig7}
\end{figure}

In the conventional manner of defining the lifetime for exponential decay,
such as is done for virtual states, the
metastable state inside the well would have an infinite lifetime. 
Although this is not a proof at all,
the numerical results strongly suggest that, if not infinite, the lifetime of
the metastable state is extremely large. 
Instead of being a mere transient effect, 
that disappears after a time comparable to the transit time of the packet 
through the well, as other transients, like the diffraction in time
process\ci{moshinsky}, it looks as if the diffraction pattern persists.
The metastable state inside the well is, in a way,
a decaying trapped state. 

\section{\sl Two dimensional wave packet scattering from an attractive well}

The results of the previous section call for a more realistic
calculation. 
As a step towards a full three dimensional calculation, we proceed here to
describe the results of the two-dimensional case.

Consider two-dimensional scattering off a potential well as described by the
time dependent Schr\"odinger equation

\be\label{sch}
\frac{-1}{2~m}\bigg(\frac{\dd^2}{\dd r^2}+\frac{\dd}{r\dd r}+
\frac{\dd^2}{r^2\dd \phi^2}\bigg)\Phi+~V(r)~\Phi=~i\frac{\dd \Phi}{\dd t}
\ee

Where $\phi$ is the polar angle and $r$, the radial coordinate.
Expanding in partial waves,

\be\label{partial}
\Phi(t,r,\phi)=~\sum_{l=0}^{lmax}~e^{i~l~\phi}~\phi_{l}(r,t)
\ee
we obtain decoupled partial wave equations (the potential is assumed
independent of $\phi$).

\be\label{schpar}
\frac{-1}{2~m}\bigg(\frac{\dd^2}{\dd r^2}+\frac{\dd}{r\dd r}-
\frac{l^2}{r^2}\bigg)\phi_l+~V(r)~\phi_l=~i\frac{\dd \phi_l}{\dd t}
\ee

A further simplification is achieved by the substitution
$\Phi_l=\frac{\tilde\Psi_l}{\sqrt{r}}$. The potential acquires an extra
term and the first derivative cancels out. Henceforth we work
with the wave function $\Psi=\Phi~\sqrt{r}$.
This substitution also allows for a simple numerical treatment.
For each partial wave we apply the method used in the one dimensional
case.\ci{gold}

We start the scattering event of a minimal uncertainty wave packet

\bea\label{packet1}
\Psi_0~=~C\sqrt{r}
~exp\bigg({i~q~(x-x_0)-\frac{(x-x_0)^2+(y-y_0)^2}{4~\Delta^2}}\bigg)
\eea

at a distance large enough to be outside the range of the potential,

\bea\label{well1}
V(r)=-A~exp\bigg({-\frac{r^2}{w^2}}\bigg)
\eea

for which we again use a Gaussian.

We present our results for different impact parameters $y_0$, for
a packet traveling initially along the negative $x$ axis towards the well,
with average speed $v=\frac{q}{m}$ as a function of angle and distance from the
location of the potential.

Figures 8-10 show the scattered waves at angles of 180$^o$, 90$^o$ and 0$^o$
respectively for initial momenta $q=1,2,3$ in inverse distance
units. The initial center of the packet is at $x_0=-10, y_0=0$. The
parameters of the well are $w=2,~V_0$=1, the width of the initial packet
$\Delta=0.5$. Throughout the calculation we limited the number of partial
waves to $l_{max}=50$. The accuracy in the expansion obtained with this
limit, was found to be better than 1\%. For large
impact parameters we increased the number of partial waves
up to $l_{max}=70$.
The wave functions are normalized to $2\pi$.
\begin{figure}
\epsffile{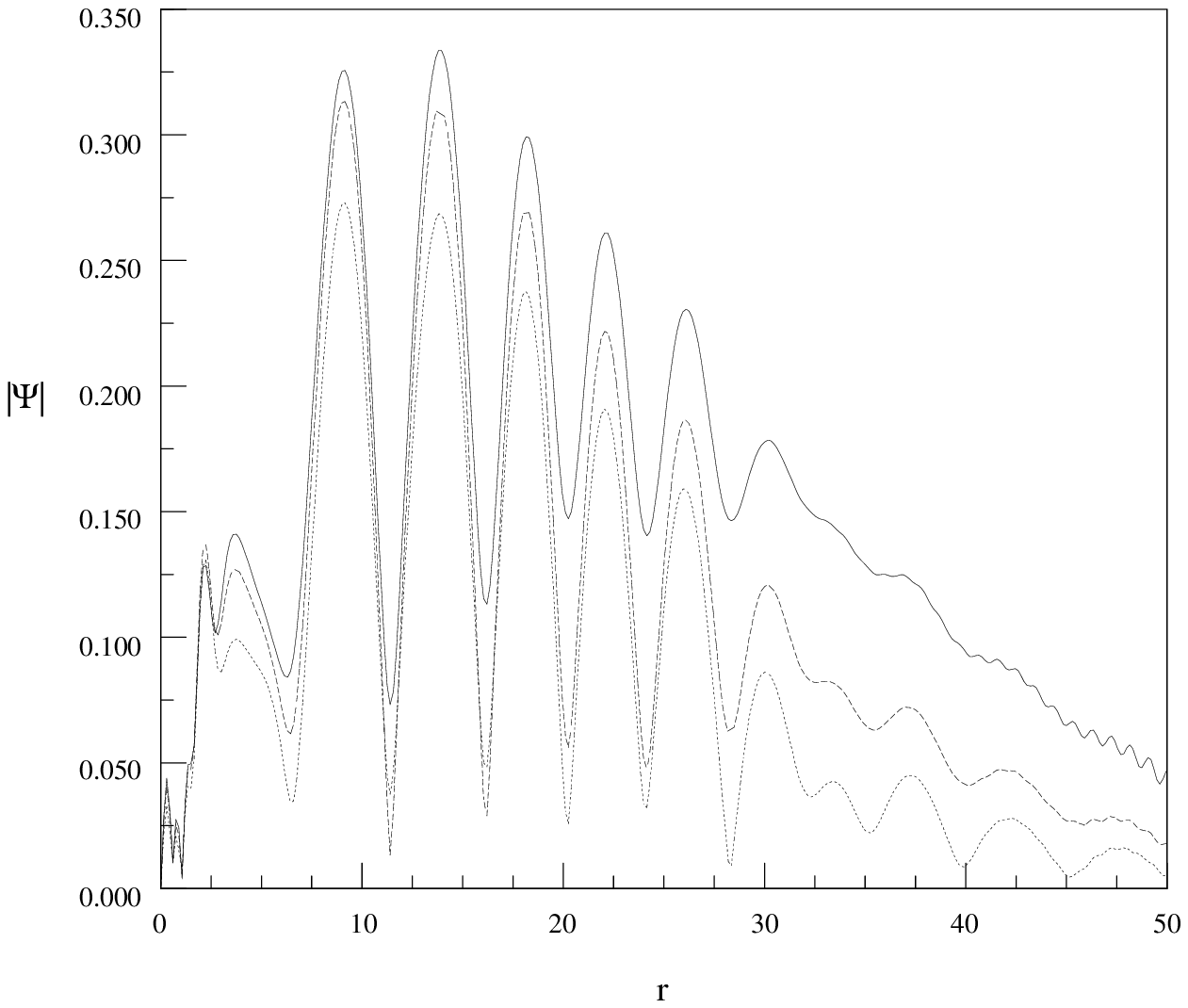}
\vsize=5 cm
\caption{\sl $|\Psi|$ at an angle of 180$^o$,
as a function of r at t=300. Wave packet width $\Delta=0.5$,
$x_0$=-10, $y_0=0$. Well width, w=2 depth $V_0$=1.
Average momenta of the packet were q=0.5 (solid line),
q=1 (dashed line), q=1.5 (small dash line)}
\label{fig8}
\end{figure}
\begin{figure}
\epsffile{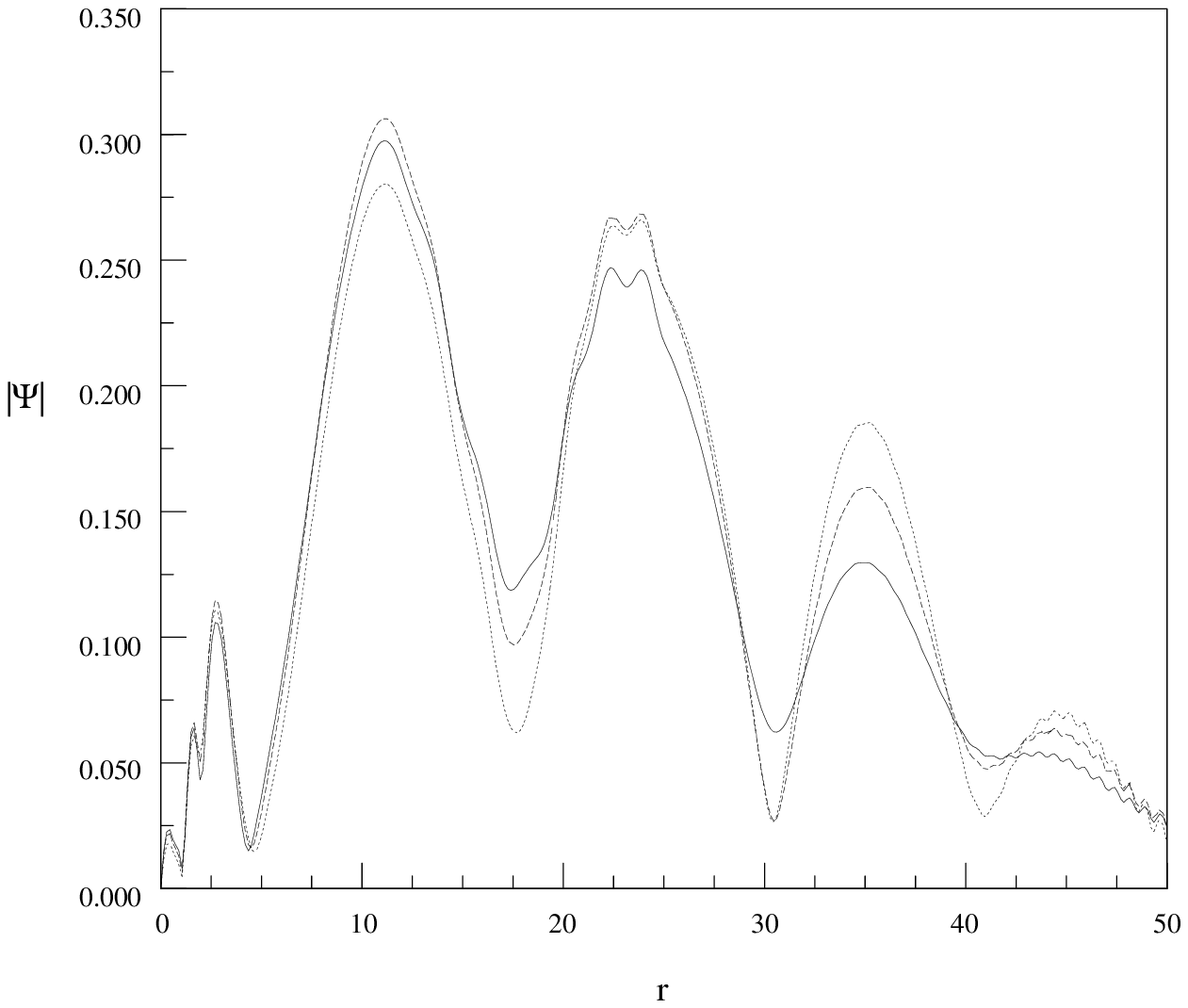}
\vsize=5 cm
\caption{\sl $|\Psi|$ at an angle of 90$^o$,
as a function of r at t=300. Wave packet width $\Delta=0.5$,
$x_0$=-10, $y_0=0$. Well width, w=2 depth $V_0$=1.
Average momenta of the packet were q=0.5 (solid line),
q=1 (dashed line), q=1.5 (small dash line)}
\label{fig9}
\end{figure}
\begin{figure}
\epsffile{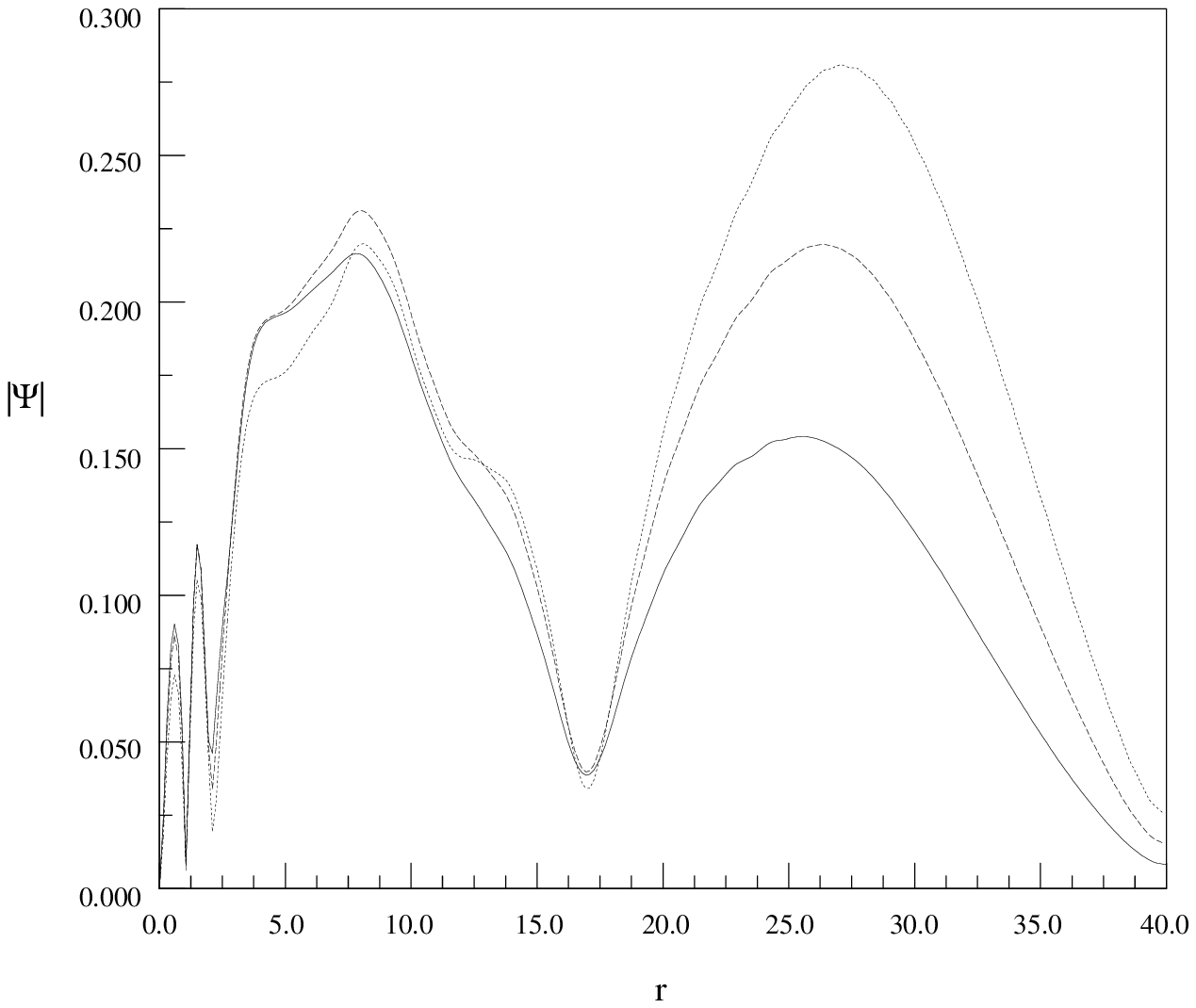}
\vsize=5 cm
\caption{\sl $|\Psi|$ at an angle of 0$^o$,
as a function of r at t=300. Wave packet width $\Delta=0.5$,
$x_0$=-10, $y_0=0$. Well width, w=2 depth $V_0$=1.
Average momenta of the packet were q=0.5 (solid line),
q=1 (dashed line), q=1.5 (small dash line)}
\label{fig10}
\end{figure}

The figures show clearly that the same phenomenon
found in the one dimensional case emerges in two dimensions.
Even at small angles the effect persists, although the multiple
peak structure is cleaner in the backward direction.
Figure 11 shows the comparison between backward and forward scattering for
impact parameter $y_0=0$. Large angle scattering
shows up as an extremely important element.

\begin{figure}
\epsffile{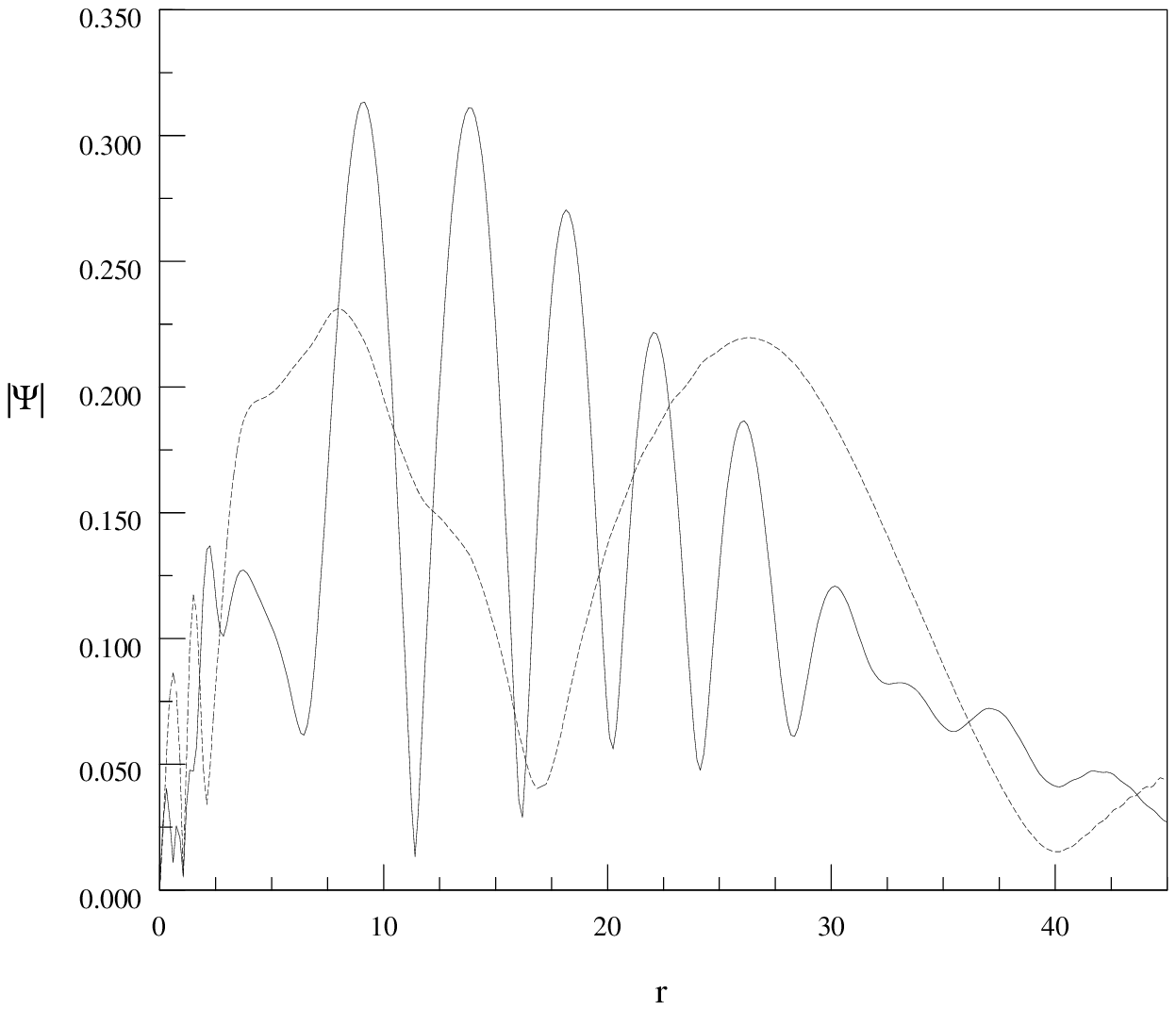}
\vsize=5 cm
\caption{\sl $|\Psi|$s at 180$^o$ (solid line), and
0$^o$ (dashed line) for a packet with q=1 and zero impact parameter at t=300.
Well parameters remain as in figure 7}
\label{fig11}
\end{figure}

Figures 12-13 show the behavior of the scattered wave at large and small angles
for increasing impact parameter.
\begin{figure}
\epsffile{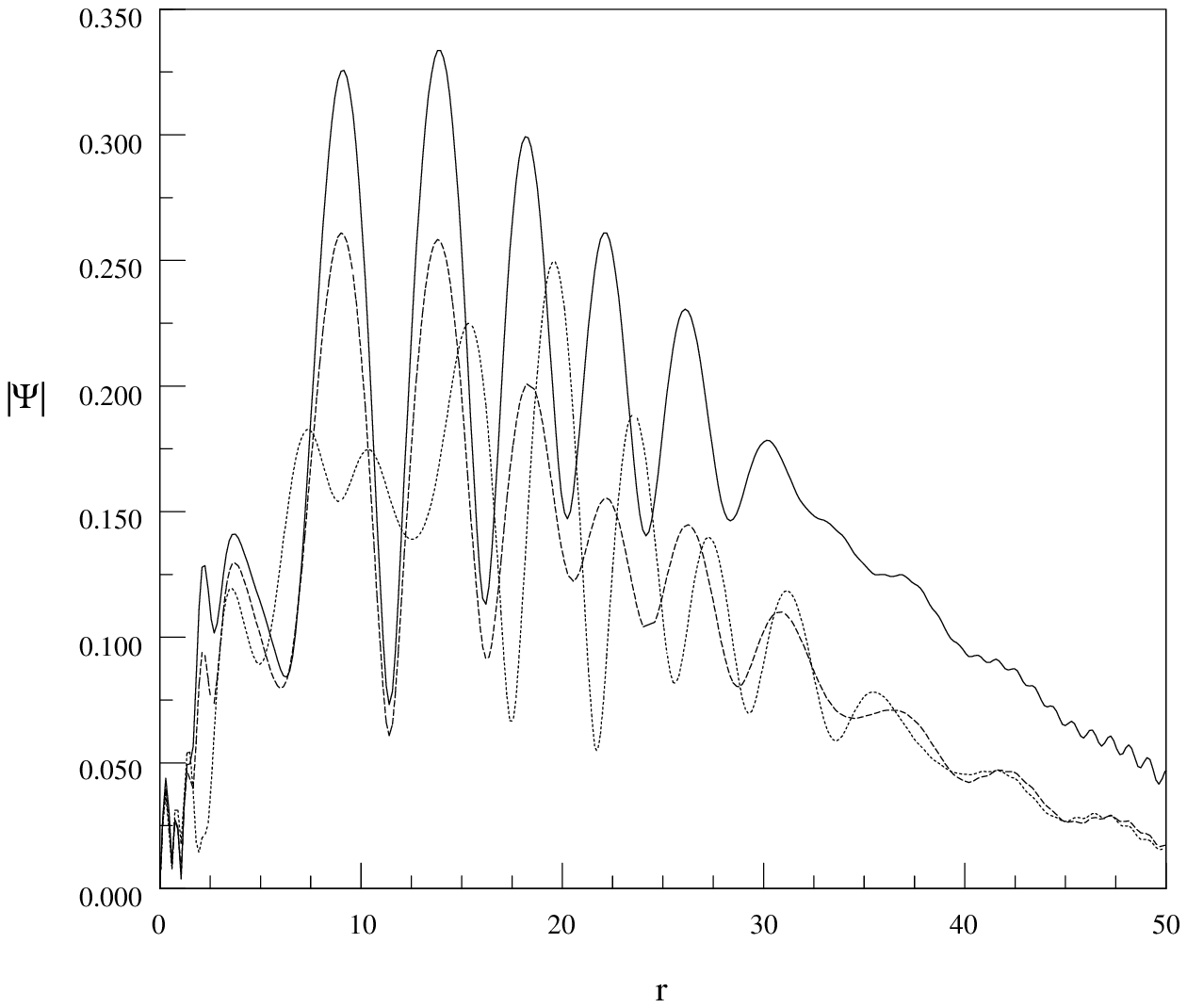}
\vsize=5 cm
\caption{\sl $|\Psi|$s at 180$^o$ for impact parameter
$y_0=0$, solid line, $y_0=1.5$, dashed line, and $y_0=3$, small dash line
at t=300.
Well and packet parameters as in figure 7}
\label{fig12}
\end{figure}
\begin{figure}
\epsffile{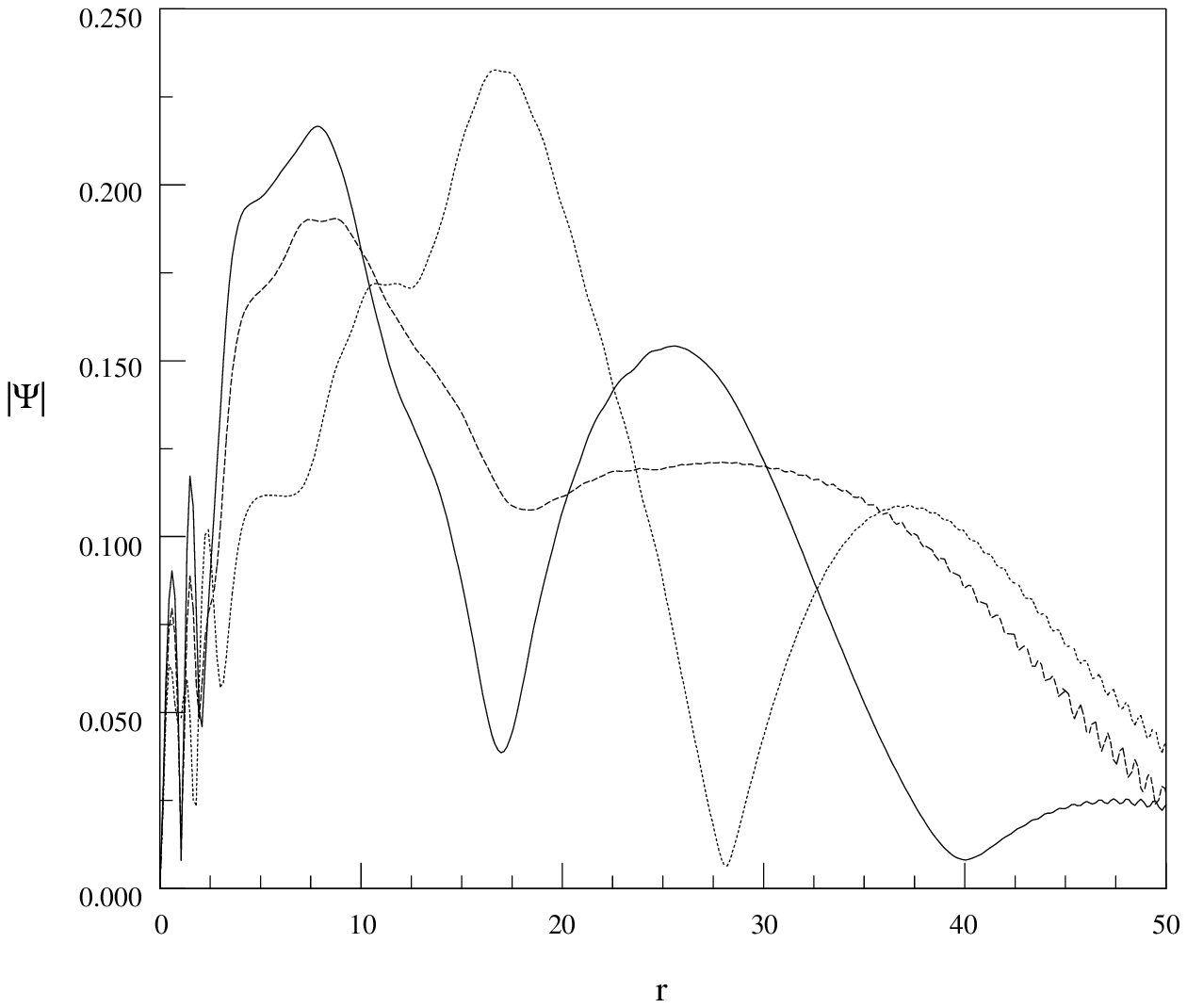}
\vsize=5 cm
\caption{\sl $|\Psi|$ at 0$^o$ for impact parameter
$y_0=0$, solid line, $y_0=1.5$, dashed line, and $y_0=3$, small dash line 
at t=300.
Well and packet parameters as in figure 8}
\label{fig13}
\end{figure}

At backward angles, the impact parameter
influences the shape of the pattern very little.
The memory of the initial information concerning impact parameter
-and even momentum, for moderate momenta as compared to the
inverse of the well width- is erased.

We can visualize the existence of a quasi-bound state inside the
well by selecting the region around the origin and plotting
real and imaginary parts of the wave function.
Figures 14 and 15 show these waves for masses $m=20,5$ for
backward angles. Here we depicted $\Phi$ instead of $\Psi$.
\begin{figure}
\epsffile{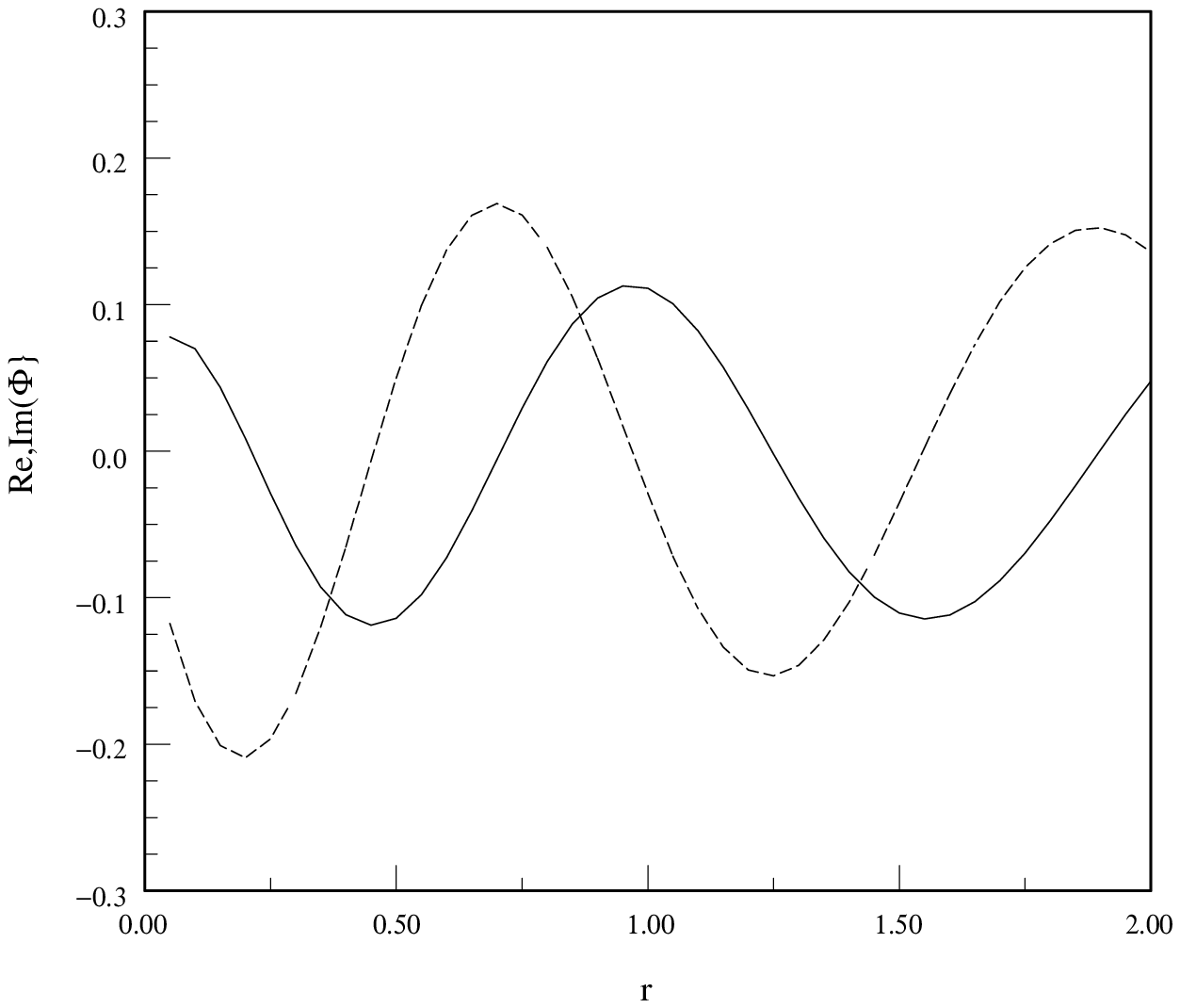}
\vsize=5 cm
\caption{\sl Real (solid line) and Imaginary (dashed line) part
of $\Phi$ at 180$^o$ for impact parameter
$y_0=0$, momentum $q=1$, mass $m=20$, in a region inside the well 
at t=150.}
\label{fig14}
\end{figure}
\begin{figure}
\epsffile{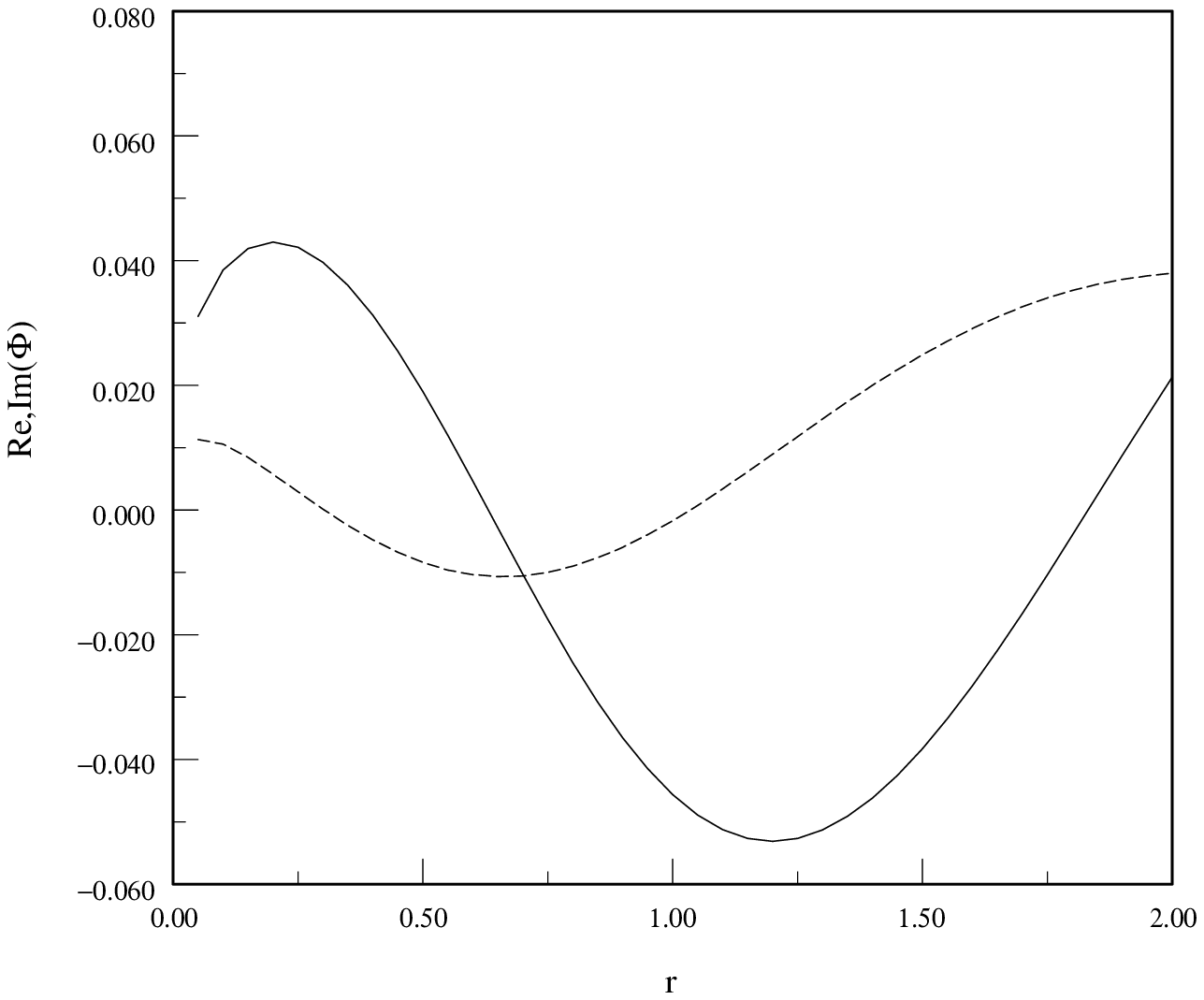}
\vsize=5 cm
\caption{\sl Real (solid line) and Imaginary (dashed line) part
of $\Phi$ at 180$^o$ for impact parameter
$y_0=0$, momentum $q=1$, mass $m=5$, in a region inside the well 
at t=150.}
\label{fig15}
\end{figure}
As for the one dimensional case, we see a sinusoidal behavior.
For $m=20$ it is seen that the well width accomodates approximately
two wavelengths, while for $m=5$ one wavelength fits in.
(Recall that the well does not have a sharp edge)
From the figures we can read off the value of the wavenumber inside the
well, namely $k'=\sqrt{k^2+2~m~|V_0|}$ to be $k'=\frac{2\pi~n}{2~w}$. 
For m=20 we find n=2, and for m=5 we find n=1. Inserting the values
of the mass and the depth of the potential $V_0$, we obtain $k\approx 0$.

\begin{figure}
\epsffile{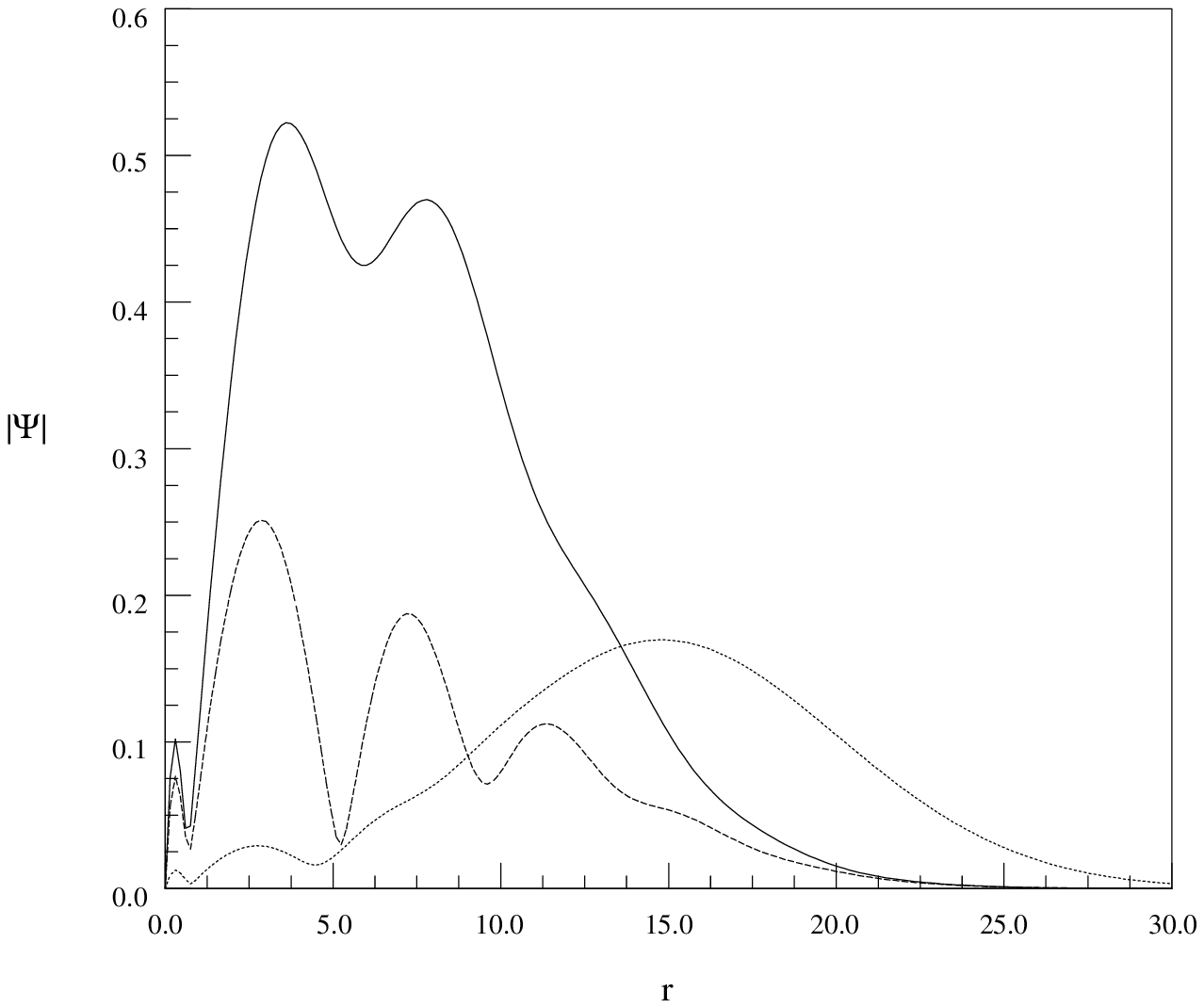}
\vsize=5 cm
\caption{\sl $|\Psi|$ at an angle of 180$^o$,
as a function of r. Wave packet width $\Delta=2$,
$x_0$=-10, $y_0=0$. Well width, w=0.5 depth $V_0$=1.
Average momenta of the packet were q=0.5 (solid line),
q=1 (dashed line), q=1.5 (small dash line) at t=300.}
\label{fig16}
\end{figure}
\begin{figure}
\epsffile{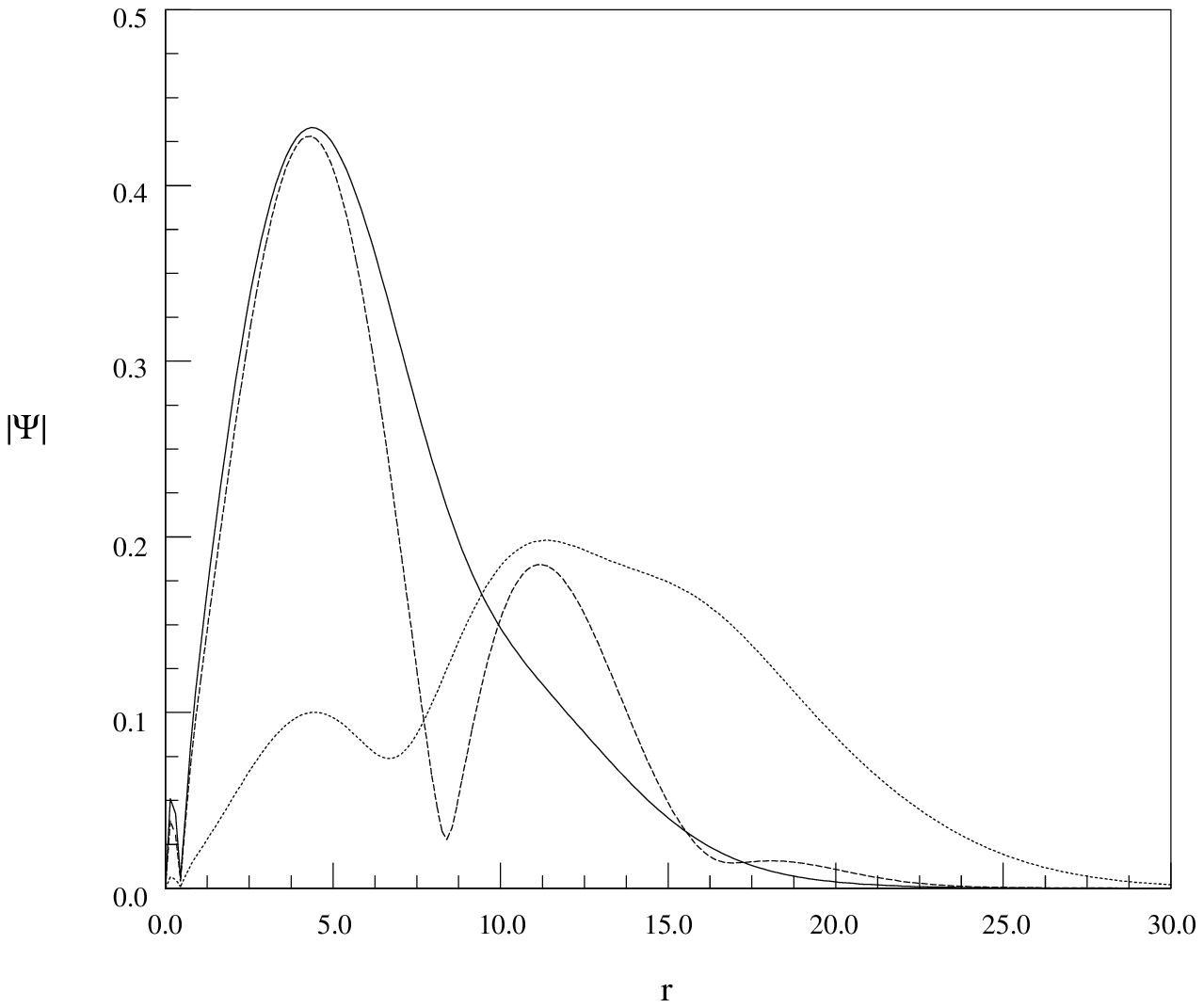}
\vsize=5 cm
\caption{\sl $|\Psi|$ at an angle of 90$^o$,
as a function of r. Wave packet width $\Delta=2$,
$x_0$=-10, $y_0=0$. Well width, w=0.5 depth $V_)$=1.
Average momenta of the packet were q=0.5 (solid line),
q=1 (dashed line), q=1.5 (small dash line) at t=300.}
\label{fig17}
\end{figure}
\begin{figure}
\epsffile{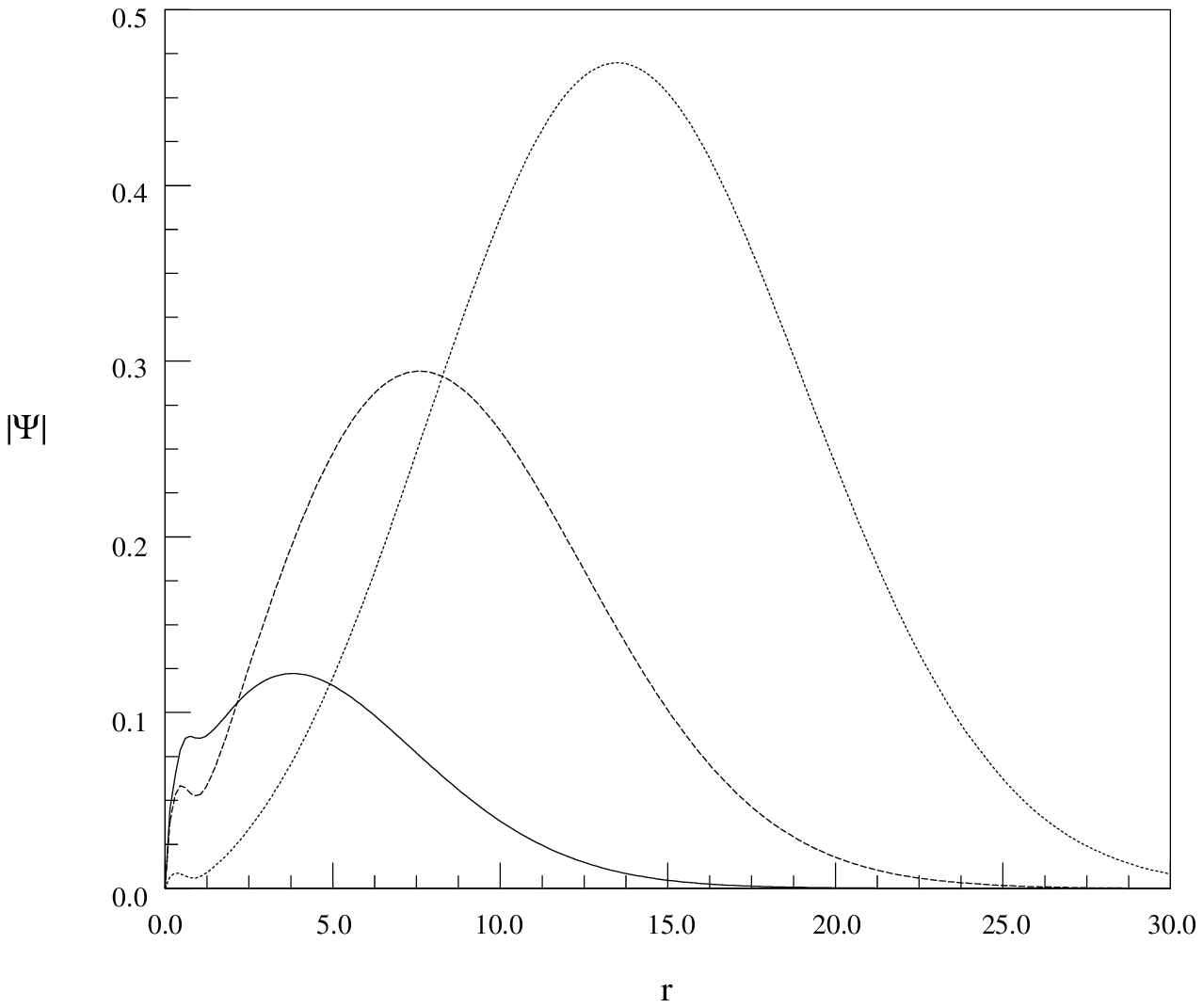}
\vsize=5 cm
\caption{\sl $|\Psi|$ at an angle of 0$^o$,
as a function of r. Wave packet width $\Delta=2$,
$x_0$=-10, $y_0=0$. Well width, w=0.5 depth $V_0$=1.
Average momenta of the packet were q=0.5 (solid line),
q=1 (dashed line), q=1.5 (small dash line) at t=300.}
\label{fig18}
\end{figure}

The two-dimensional case resembles remarkably the
one dimensional scattering, for packets that are initially narrower than the
well. This is also true for the scattering of wide packets.
Figures 16-18 show that all the polychotomous wave trains disappear when
the initial packet is wider than the well.
The multiple peaks are now absent.
Forward and backward scatterings look now quite similar.

The effect depends on the existence of a quasi-bound state inside the
well. Shallow potentials cannot sustain the metastable states.
Therefore, the polychotomous behavior should gradually disappear when
the depth of the well is diminished.
Figure 19 shows one such case for a well amplitude of $A=0.03$
\begin{figure}
\epsffile{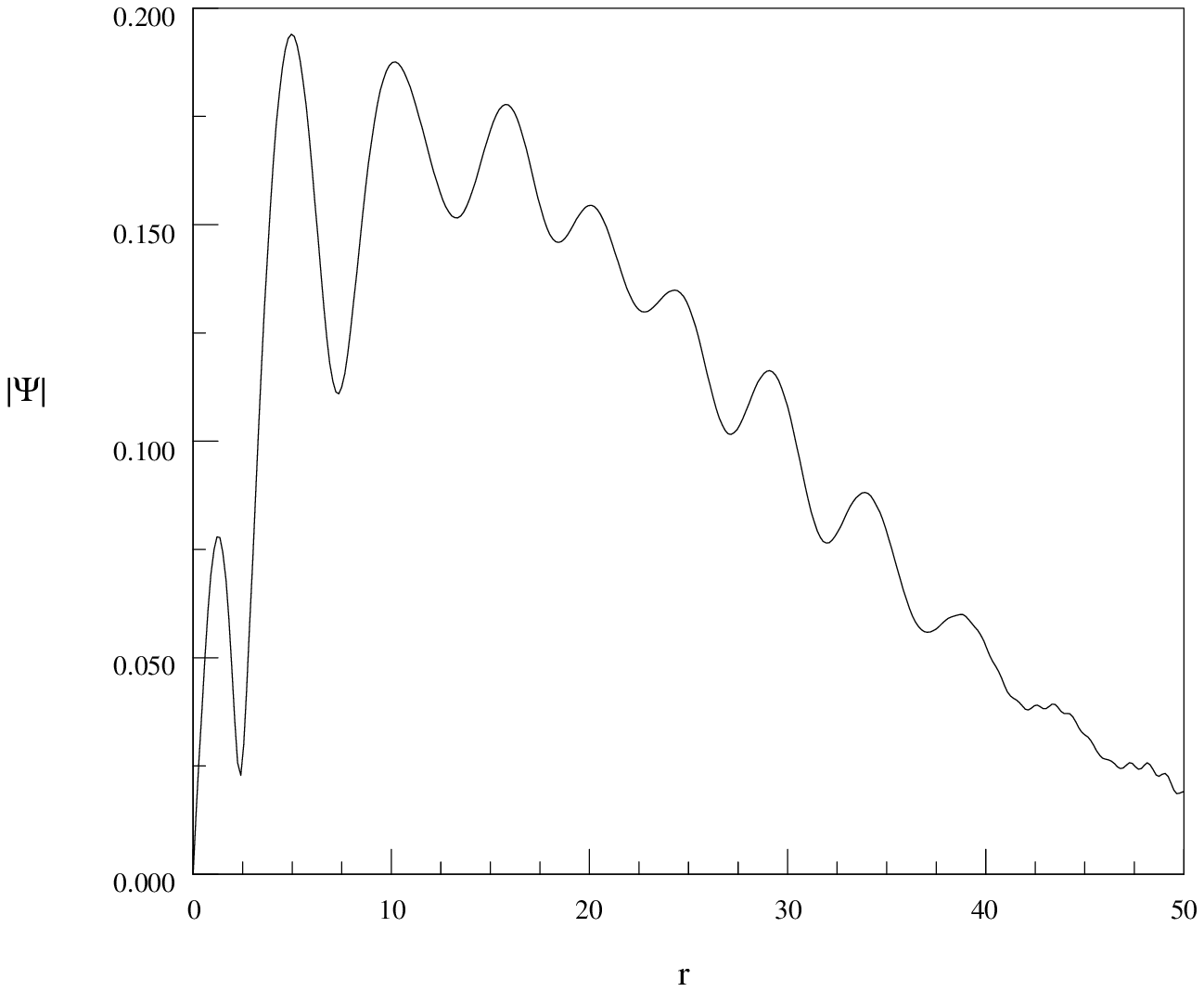}
\vsize=5 cm
\caption{\sl $|\Psi|$ for a scattering angle of 180$^0$,
packet width $\Delta=0.5$ starting at $x_0$=-10 impinging upon a well
of width parameter w=2 and depth $V_0$=0.03 at t=300, the initial
average momentum of the packet is q=1, with a mass of $m=20$.}
\label{fig19}
\end{figure}

\newpage
\section{\sl Suggested experiments}

We have found that the polychotomous coherent effect of
ref.\ci{pra}, persists in two dimensions and presumably this will
be true in a full three dimensional calculation.

Experimental work may take advantage of these findings and design
setups to research the phenomena described here.
We mentioned in section 1 the possibility of using
cavity experiments with atomic beams. The assessment of feasibility
of such experiments is however, beyond the knowledge of the author. Although,
it appears, a tangible option.

Experiments in optics, sound propagation, microwaves, might also
be appropriate in order to find the polychotomous behavior.

Another alternative that seems viable,
consists in an experiment related to those known under
the title of {\sl ALAS}.\ci{michel}

{\sl ALAS} stands for anomalous large angle scattering.
It occurs for $\alpha$ scattering on certain closed shell nuclei for
incident energies below 100 MeV.
The backward scattering is so pronounced that can exceed the Rutherford cross
section by several orders of magnitude.
Although many explanations based on optical models have been provided over the
years for this process, it remains rather obscure.
A possible interpretation based on the present results would be
that, the $\alpha$ particle is a finite extent system of dimensions
smaller than the nuclear well. Considered as a wave packet
it could form a metastable state inside the
well in a similar manner to the packets dealt with presently.
The large backward scattering is then a reflection of the behavior
found here for a finite size packet.
A clear imprint of the effect would however, require
the detection of the $\alpha$ particles as a function of time
in order to observe the oscillatory amplitudes that dominate
at large scattering angles.
Data acquisition in nuclear (and other) experiments generally
averages over time variations, except for coincidence
experiments. The multiple peak behavior demands a continuous
time dependent recording of the alpha particles, triggered
by the bunches emitted from the accelerating machine.
If experimental support is indeed gathered, then the
effect can be turned around in order to become a
research tool, due to its dependence on the geometrical and
dynamical parameters of both projectiles and target.
A firm theoretical connection to the {\sl ALAS} effect,
requires, eventually, a much more laborious theoretical
and numerical work than the one carried out here.
Efforts in that direction are currently underway.

{\bf Acknowledgements}

This work was supported in part by the Department of
Energy under grant DE-FG03-93ER40773 and by the National Science Foundation
under grant PHY-9413872, while the author was on sabbatical
at the cyclotron institute of the Texas A\&M University.
It is a pleasure to thank Prof. Youssuf El Masri of the UCL, University
of Lovain-la-Neuve, Belgium for the information concerning the ALAS
effect.

\end{document}